\documentclass[fontsize=9pt, DIV=calc, a4paper, twocolumn, headings=standardclasses]{scrartcl}

\usepackage[utf8]{inputenc}
\usepackage{lmodern}
\usepackage[english]{babel}
\usepackage[activate={true, nocompatibility}, final, tracking=true, kerning=true, factor=1100, stretch=10, shrink=10]{microtype}
\usepackage{booktabs, graphicx, amsmath, amssymb, physics, mathtools, bm}
\usepackage[format=plain, labelfont=bf]{caption}
\usepackage[auth-lg, affil-it]{authblk}
\usepackage[sort&compress, numbers]{natbib}
\usepackage[frozencache,cachedir=minted-cache]{minted}
\usepackage[hidelinks]{hyperref}

\makeatletter
\doforeachtocfile{\unsettoc{\@currext}{onecolumn}}
\makeatother

\renewcommand{\vec}{\bm}

\newcommand{\program}{TinyDEM}
\title{\program{}: Minimal open granular DEM code\\with sliding, rolling and twisting friction}

\author{Roman Vetter}

\affil{Department of Biosystems Science and Engineering, ETH Z\"{u}rich, Schanzenstrasse 44, 4056 Basel, Switzerland}

\date{\today}

\begin{document}

\maketitle

\noindent\textbf{This article introduces \program{}, a lightweight implementation of a full-fledged discrete element method (DEM) solver in 3D. Newton's damped equations of motion are solved explicitly for translations and rotations of a polydisperse ensemble of dry, soft, granular spherical particles, using quaternions to represent their orientation in space without gimbal lock. Particle collisions are modeled as inelastic and frictional, including full exchange of torque. With a general particle-mesh collision routine, complex rigid geometries can be simulated. \program{} is designed to be a compact standalone program written in simple C++11, devoid of explicit pointer arithmetics and advanced concepts such as manual memory management or polymorphism. It is parallelized with OpenMP and published freely under the 3-clause BSD license. \program{} can serve as an entry point into classical DEM simulations or as a foundation for more complex models of particle dynamics.}

\vskip\baselineskip

\noindent\textbf{$^*$Correspondence:} vetterro@ethz.ch

\tableofcontents

\section{Introduction}

This paper does not introduce new concepts or methods. It does not solve a long-standing or particularly challenging problem, nor does it allow to simulate something that was impossible before. It is not concerned with relativistic or quantum physics. Instead, this article has something to offer that is rarely found in published computer programs: It introduces a lean, easy-to-read standalone implementation of an elementary ingredient for granular (macroscopic) particle simulations, named \program{}. \program{} is unique not in \emph{what} it does, but in \emph{how} it does it. If you are a student or developer of own simulations of particle dynamics, looking for a simple DEM program to use, or just interested in a neat coding example, read on.

The discrete (or disjunct) element method (DEM) was pioneered by Cundall and Strack who developed the first 2D implementation named BALL in the 1970s \cite{Cundall:1979}. In the DEM, a granular medium is represented by an ensemble of spherical or non-spherical particles that move, rotate and collide with each other, exchanging forces and torques in the process. The first open 3D granular DEM implementation with a source code that is still available today appears to be YADE-DEM from 2008, which is integrated in the multi-purpose physics simulation framework YADE \cite{Kozicki:2008}. Numerous other codes with varying degrees of complexity and capability followed, published mostly under a GNU General Public or BSD license. They are reviewed in Table~\ref{tab:1}. It is fair to say that a good selection of powerful open-source DEM programs are available nowadays, at least for a technically versed audience. Nine popular ones of them were recently compared and benchmarked \cite{Dosta:2024}, revealing largely good agreement between model predictions, despite occasionally varying implementation details.

\begin{figure}
	\centering
	\includegraphics[width=0.9\linewidth]{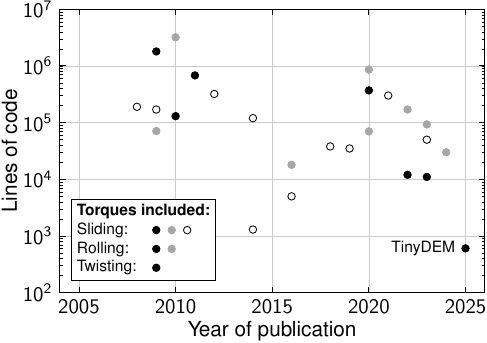}
	\caption{\textbf{Volume and level of detail of the contact model of published open-source soft-sphere DEM codes.} Code volume is measured in number of lines as detailed in Table~\ref{tab:1}, and includes features beyond the DEM part, where applicable. Codes represented by a filled dot have all three modes of torque included in their friction model, gray dots lack the twisting friction, and empty dots lack both rolling and twisting friction.
	}
	\label{fig:1}
\end{figure}

Why introduce yet another one, then, one might ask. From the code volume quantification in Table~\ref{tab:1}, it becomes evident that, although writing good and versatile DEM code is not rocket science, it is usually still a considerable endeavor. There is no complete classical DEM code available with under 10,000 lines of code, which is easily doable in principle, as will be shown here. Fig.~\ref{fig:1} shows this graphically: Published DEM frameworks often exceed 100,000 lines of code (including whitespace and comments), even when the friction model used for particle collisions is incomplete. While this is certainly in part because DEM code is often embedded in larger multiphysics frameworks whose functionality far exceeds that of just the DEM part, it sets the hurdle to start new computational DEM projects unnecessarily high. Large frameworks can be difficult to install, operate, edit, and maintain. Only two open DEM codes (ppohDEM and Blaze-DEM) have a code volume lower than 10,000 lines.

Fig.~\ref{fig:1} highlights an additional limitation of the current body of open-source DEM frameworks: Most of them model frictional collisions only partially. Appropriate representation of particle rotations and friction is important in crack propagation \cite{Wang:2008}, soil mechanics simulations \cite{Oquendo:2011}, mixing of grains \cite{Remy:2009}, collapse of granular assemblies \cite{Suiker:2004}, and to measure stress–strain relations of granular materials \cite{Qian:2023}. Out of 24 reviewed codes, only six model all three modes of frictional torque exchange: sliding, rolling, and twisting. Ten include only sliding friction, some of them limited to history-independent kinetic friction, making them unsuited for the simulation of particle configurations in static equilibrium. Both codes below 10,000 lines lack rolling and twisting friction models. 

There is clear value in simple standalone code---be it for education, method development, or other purposes, especially when code reusability is a concern. Several other lightweight programs that focus on doing one thing very well have demonstrated this paradigm: TinyFSM \cite{Burri}, SVL \cite{Wilmott}, jsmn \cite{Zaitsev}, TinyXML-2 \cite{Thomason}, tinyMD \cite{Machado:2021}, MMM1D \cite{Vetter:2019}, PolyHoop \cite{Vetter:2024}, CImg \cite{Tschumperle:2012}, and the suite of mini-apps contained in the Mantevo project \cite{Heroux:2009} are examples of minimalistic code that can be useful in many situations, from small student projects to large software libraries. They are often published under permissive licenses, which enhances their usefulness.

In this spirit, this work introduces \program{}, an exceptionally compact, portable and easy-to-use standalone DEM simulator that includes all three modes of frictional torque exchange, and an accurate representation of particle orientation with quaternions. \program{} is released under the BSD 3-clause license that permits both commercial and non-commercial use. It consists of only about 600 lines of commented C++11 code, has no dependencies, and is parallelized with OpenMP, requiring no access to high-performance computer networks or GPUs. The source code is designed not only for compactness and readability, but also to be as direct and instructive as possible, containing no magical numbers, numerical tolerances, explicit pointer arithmetics, manual memory management or polymorphism beyond what the standard template library provides. A unified collision routine handles both particle-particle and particle-mesh contacts with the same general code.

This article serves as \program{}'s comprehensive documentation: All essential program features are described in the text, and in turn, all physics described here can be found in this form in the source code. After a detailed model description, some classical example simulations are showcased, and instructions for installation, usage and postprocessing are given. Finally, a computational benchmark is provided as a basis for runtime and memory requirement estimation in the user's specific scenario.

\begin{table*}
\centering
\small
\caption{\textbf{Overview of open-source DEM programs.} Only models that include frictional collisions of soft spherical granular particles in 3D are listed. Number of code lines, where available, are rounded and based on latest available versions at the time of research (January 2025), including comments, whitespace, and supplied dependencies, but excluding dependencies that are not supplied with the code. Where available, the year of publication of the accompanying paper is used; otherwise, the approximate introduction year of the core DEM functionality is taken. OS: Open source; HM: Hertz--Mindlin; DMT: Derjaguin--Muller--Toporov; JKR: Johnson--Kendall--Roberts; HKK: Hertz--Kuwabara--Kono.}
\label{tab:1}
\setlength{\tabcolsep}{0.3em}
\begin{tabular}{@{}p{0.6cm}lp{4.5cm}p{5.9cm}p{1.4cm}p{1.45cm}p{0.95cm}@{}}
\toprule
\textbf{Year} & \textbf{Name} & \textbf{Contact model} & \textbf{Implementation} & \textbf{OS} & \textbf{License} & \textbf{Ref.}\\
\midrule
2005 & --- & Hooke/HM/HKK, static \& kinetic Coulomb sliding & Modified MD code ``DL\_POLY\_2'' in Fortran 90, MPI & partial (printed) & \,--- & \cite{Dutt:2005}\\\midrule
2008 & YADE & HM, Mohr--Coulomb sliding & 190,000 lines of C++/Python, OpenMP, several dependencies (boost, eigen, Qt, freeglut3 etc.) & yes & GPL 2 & \cite{Kozicki:2008}\\\midrule
2009 & ESyS-Particle & Hooke/HM, static \& kinetic Coulomb sliding & 170,000 lines of C++/Python, MPI, some dependencies (boost) & yes & Apache 2.0 & \cite{Wang:2009}\\\midrule
2009 & LAMMPS & Hooke/HM/DMT/JKR, static \& kinetic Coulomb friction (sliding, rolling \& twisting) & 1.8 mio.\ lines of C++, MPI & yes & GPL 2 & \cite{Kloss:2010}\\\midrule
2009 & MechSys & Hooke, static \& kinetic Coulomb friction (sliding, rolling) & 71,000 lines of C++, OpenMP, CUDA, several dependencies (boost, GSL, LAPACK etc.) & yes & GPL 3 & \cite{Galindo-Torres:2009}\\\midrule
2010 & Woo & Various, static \& kinetic Coulomb friction (sliding, rolling \& twisting) & Fork of YADE, 130,000 lines of C++/Python, OpenMP, several dependencies (boost, eigen etc.) & yes & GPL 2 & \cite{Smilauer:2010}\\\midrule
2010 & Kratos & Various, static \& kinetic Coulomb sliding, kinetic rolling & 3.2 mio.\ lines of C++/Python, OpenMP, MPI, dependency on boost & yes & BSD-3 & \cite{Dadvand:2010}\\\midrule
2011 & LIGGGHTS & Hooke/HM, static \& kinetic Coulomb sliding \& rolling & 680,000 lines of C++, MPI & yes & GPL 2 & \cite{Kloss:2011}\\\midrule
2012 & MFIX-DEM & Hooke/HM, static \& kinetic Coulomb sliding & 320,000 lines of Fortran 90/Python (Conda), OpenMP; numerous dependencies & registration \& approval required & \,--- & \cite{Garg:2012}\\\midrule
2014 & ppohDEM & Hooke, static \& kinetic Coulomb sliding & Single file, 1300 lines of Fortran 90, OpenMP, MPI & yes & CPC non-profit & \cite{Nishiura:2014}\\\midrule
2014 & GranOO & Hooke, kinetic Coulomb sliding & 120,000 lines of C++/Python, several dependencies (boost, zlib, eigen etc.) & yes & GPL 3 & \cite{Andre:2014}\\\midrule
2016 & Blaze-DEM(GPU) & Hooke, kinetic Coulomb sliding & 5000 lines of C++, CUDA & yes & BSD-3 & \cite{Govender:2015, Govender:2016}\\\midrule
2016 & cemfDEM & HM, static \& kinetic Coulomb sliding, kinetic rolling & 18,000 lines of Fortran 90 & yes & GPL 3 & \cite{Norouzi:2016}\\\midrule
2018 & ParaEllip3d & HM, static \& kinetic Coulomb sliding & 38,000 lines of C++, MPI, dependency on boost, eigen, qhull & yes & MIT, Apache 2.0 & \cite{Yan:2018}\\\midrule
2019 & OpenFPM & Hertz, static \& kinetic Coulomb sliding & Reimplementation of \cite{Walther:2009} in 35,000 lines of C++, CPU \& GPU parallelization, several dependencies (boost, OpenMPI, libhilbert etc.) & yes & BSD-3 & \cite{Incardona:2019}\\\midrule
2020 & MercuryDPM & Various, static \& kinetic Coulomb friction (sliding, rolling \& twisting) & 370,000 lines of C++14, Fortran \& Python, MPI & yes & BSD-3 & \cite{Weinhart:2020}\\\midrule
2020 & MUSEN & Various, static \& kinetic Coulomb sliding, kinetic rolling & 70,000 lines of C++, CUDA, dependency on Qt, protobuf, zlib & yes & BSD-3 & \cite{Dosta:2020}\\\midrule
2020 & Chrono::Granular & Various, static \& kinetic Coulomb sliding, kinetic rolling & 860,000 lines of C++, CUDA & yes & BSD-3 & \cite{Kelly:2020, Fang:2021}\\\midrule
2021 & SudoDEM & Identical to YADE & Derived from YADE (same dependencies), 300,000 lines of C++, Python & yes & GPL 3 & \cite{Zhao:2021}\\\midrule
2022 & Lethe-DEM & Hooke/HM/JKR/DMT, static \& kinetic Coulomb sliding, kinetic rolling & 170,000 lines of C++/Python, MPI, several dependencies (deal.II, numdiff, p4est, trilinos, METIS) & yes & Apache 2.0, LGPL 2.1 & \cite{Golshan:2023}\\\midrule
2022 & DEMBody & HM, static \& kinetic Coulomb friction (sliding, rolling \& twisting) & 12,000 lines of Fortran 90, OpenMP & registration \& approval required & non-commercial agreement & \cite{Cheng:2022}\\\midrule
2023 & PhasicFlow & Hooke/HM/HKK, static \& kinetic Coulomb sliding, kinetic rolling & 93,000 lines of C++, OpenMP, CUDA, some dependencies (Kokkos, tbb) & yes & GPL 3 & \cite{Norouzi:2023}\\\midrule
2023 & --- & JKR, static \& kinetic Coulomb friction (sliding, rolling \& twisting) & Builds on MechSys; 11,000 lines of C++, OpenMP & yes & GPL 3 & \cite{Qian:2023}\\\midrule
2023 & CP3d & Hooke/HM, static \& kinetic Coulomb sliding & 50,000 lines of Fortran 95, MPI, dependency on FFTW & yes & MIT & \cite{Gong:2023}\\\midrule
2024 & DEM-Engine & HM, static \& kinetic Coulomb sliding, kinetic rolling & Builds on Chrono; 30,000 lines of C++ (not counting Chrono), CUDA & yes & BSD-3 & \cite{Zhang:2024}\\
\bottomrule
\end{tabular}
\end{table*}

\section{Particle model}

\subsection{Equations of motion}

Like many other soft-sphere DEM codes, \program{} solves the damped Newton--Euler equations for translation and rotation. Each particle $i=1,...,N$ in the ensemble carries its own position vector $\vec{x}_i$, velocity $\vec{v}_i$, orientation $\vec{q}_i$ (discussed later), angular velocity $\vec{\omega}_i$, and radius $R_i$. Bold symbols represent vectorial quantities. Particles move according to \cite{Remy:2009}
\begin{align}
	m_i\frac{d\vec{v}_i}{dt} &= \sum_{j\neq i}\left(\vec{f}_\mathrm{n}^{ij}+\vec{f}_\mathrm{s}^{ij}\right) + m_i \vec{g} - c_im_i\vec{v}_i\label{eq:translation}\\
	I_i\frac{d\vec{\omega}_i}{dt} &= \sum_{j\neq i}\left(\vec{\tau}_\mathrm{s}^{ij} + \vec{\tau}_\mathrm{r}^{ij} + \vec{\tau}_\mathrm{t}^{ij}\right) - \gamma_i I_i\vec{\omega}_i\label{eq:rotation}
\end{align}
where $m_i=\frac{4}{3}\pi \rho_i R_i^3$ is the particle mass, $I_i=\frac{2}{5}m_i R_i^2$ the moment of inertia. In general, the vector $\vec{g}$ can be any external acceleration field, but typically it represents the gravitational acceleration.

The last terms in Eqs.~\ref{eq:translation} and \ref{eq:rotation} are often used to enhance numerical stability or to introduce energy dissipation that slows down particle motion to drive the ensemble toward a static equilibrium with linear and angular damping rates $c_i$ and $\gamma_i$. A convenient way to avoid having to tune these parameters individually is to assume drag in a (real or virtual) viscous fluid at low Reynolds numbers. Stokes's law \cite{Landau:1987} then relates both of them to a single material parameter, the dynamic viscosity $\eta$ of the fluid:
\begin{align}
	c_i &= \frac{6\pi\eta R_i}{m_i}=\frac{9\eta}{2\rho_i R_i^2},\\
	\gamma_i &= \frac{8\pi\eta R_i^3}{I_i} = \frac{10}{3}c_i.
\end{align}

\begin{figure}
	\centering
	\includegraphics[width=\linewidth]{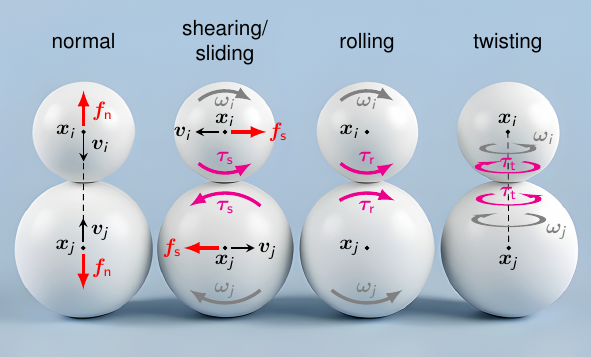}
	\caption{\textbf{The four particle-particle contact modes.} Black arrows indicate linear motion, gray arrows indicate rotation, red arrows indicate resulting forces, magenta arrows indicate resulting torques. Directions are exemplary and can also be reversed.
	}
	\label{fig:2}
\end{figure}

The sums in Eqs.~\ref{eq:translation} and \ref{eq:rotation} run over all other objects $j$ that particle $i$ interacts with. These include other particles, but also other geometric entities that may act as confining boundaries or obstacles. How nearby objects are found efficiently, such that this sum does not translate to a loop over all objects in the code, will be discussed further below. $\vec{f}^{ij}$ are the interaction forces and $\vec{\tau}^{ij}$ the torques exchanged by a pair of colliding objects. Subscript letters denote the mode or directionality of this interaction: normal (n), shearing or sliding (s), rolling (r) and twisting (t) (Fig.~\ref{fig:2}).

\subsection{Inelastic contact model}

To model the forces of dry and non-adhesive contact between elastic bodies, one needs to pick one of essentially two types of models: A simple choice is to use linear spring forces (Hooke's law), but a more realistic option is to make the repulsive force nonlinear in the indentation depth $\delta$ (as in Hertzian contact). This choice affects the behavior of particle ensembles qualitatively \cite{OHern:2003}, and most available DEM codes implement at least one of the two (Table~\ref{tab:1}). For Hertzian contact, Mindlin derived the tangential (shear) forces \cite{Mindlin:1949, Mindlin:1953}, and the combination of both became known as the Hertz--Mindlin model. \program{} implements the visco-elastic Hertz--Mindlin contact model with Coulomb friction.

In the following, the expressions for the interaction forces and torques are provided with the particle pair superscript $ij$ omitted for better readability. The sign convention is such that vectorial quantities refer to the property (position, velocity, etc.) of body $i$ relative to that of body $j$, or an effect (force, torque) on body $i$ due to interaction with body $j$. In brief, the normal and shear forces are computed as
\begin{align}
	&\vec{f}_\mathrm{n} = -k_\mathrm{n} \vec{u}_\mathrm{n} - c_\mathrm{n} \vec{v}_\mathrm{n},\label{eq:fn}\\
	&\vec{f}_\mathrm{s} = -k_\mathrm{s} \vec{u}_\mathrm{s} - c_\mathrm{s} \vec{v}_\mathrm{s},\label{eq:fs}\\
	&\delta=R_i+R_j-d\geq0,\\
	&\vec{d} = \vec{x}_i-\vec{x}_j,\quad d = \norm{\vec{d}},\\
	&\vec{\hat{n}} = \frac{\vec{d}}{d},\\
	&\vec{u}_\mathrm{n} = -\delta\vec{\hat{n}},
\end{align}
where $\vec{u}_\mathrm{n}$, $\vec{v}_\mathrm{n}$ are the vectors of relative normal position and velocity, and $\vec{u}_\mathrm{s}$, $\vec{v}_\mathrm{s}$ are the integrated displacement of static shearing friction and shear velocity, respectively, as defined further below. To satisfy Coulomb's friction law, the shearing force vector (Eq.~\ref{eq:fs}) is clipped to
\begin{equation}
\label{eq:Coulombs}
	\norm{\vec{f}_\mathrm{s}} \leq \mu_\mathrm{s}\norm{\vec{f}_\mathrm{n}}
\end{equation}
for a specified sliding friction coefficient $\mu_\mathrm{s}$.

\begin{figure}
	\centering
	\includegraphics[width=0.85\linewidth]{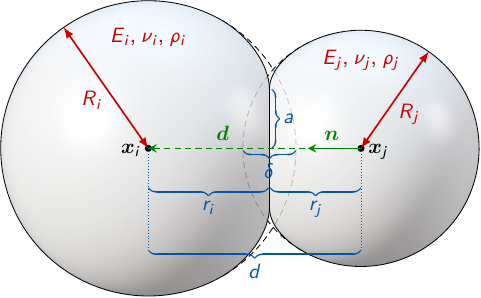}
	\caption{\textbf{Particle-particle contact.} Fundamental particle properties in red, geometric contact parameters (distances) in blue, normal vectors in green.
	}
	\label{fig:3}
\end{figure}

For two linearly elastic spherical particles, or a sphere and a half-space, the Hertz--Mindlin model provides the normal and shear stiffness coefficients as
\begin{align}
	k_\mathrm{n} &= \frac{4}{3}E^*a,\\
	k_\mathrm{s} &= 8G^*a,\label{eq:ks}\\
	a &= \sqrt{R^*\delta}
\end{align}
with effective Young's modulus, shear modulus, and radius
\begin{align}
	E^* &= \left(\frac{1-\nu_i^2}{E_i}+\frac{1-\nu_j^2}{E_j}\right)^{-1},\\
	G^* &= \left(\frac{2-\nu_i}{G_i}+\frac{2-\nu_j}{G_j}\right)^{-1},\quad G_i = \frac{E_i}{2(1+\nu_i)},\\
	R^* &= \left(\frac{1}{R_i}+\frac{1}{R_j}\right)^{-1}.
\end{align}
Note the depth-dependency of the spring coefficients via $a$, hence the nonlinearity of the contact law. $a$ is the radius of the contact area between two Hertzian spheres, which is not to be confused with the radius of overlap between two rigid spheres (Fig.~\ref{fig:3}). Since real collisions are usually inelastic, damping terms are added in Eqs.~\ref{eq:fn} and \ref{eq:fs}. The normal and shear coefficients are given by
\begin{align}
	c_\mathrm{n} &= \sqrt{5m^*k_\mathrm{n}}\,\beta,\label{eq:cn}\\
	c_\mathrm{s} &= \sqrt{\frac{1}{6}m^*k_\mathrm{s}}\,\beta,\label{eq:cs}.
\end{align}
with effective mass and damping factor
\begin{align}
	m^* &= \left(\frac{1}{m_i}+\frac{1}{m_j}\right)^{-1},\\
	\beta &= \frac{1}{\sqrt{1+(\pi/\ln e_\mathrm{n})^2}}.
\end{align}
The factor $\sqrt{5}$ in Eq.~\ref{eq:cn} originates from for the exponent $3/2$ in the distance dependency of the Hertzian normal force \cite{Antypov:2011}. For Hookean contact, that factor would be $2$ \cite{Jakubowski:1964}. $e_\mathrm{n}$ is the coefficient of normal restitution, defined as the normal velocity ratio after (') and before contact ($\vec{v}_\mathrm{n}'=-e_\mathrm{n}\vec{v}_\mathrm{n}$ with $0\leq e_\mathrm{n}\leq1$). A coefficient of tangential restitution $e_\mathrm{s}$ can be defined analogously ($\vec{v}_\mathrm{s}'=e_\mathrm{s}\vec{v}_\mathrm{s}$, with $-1\leq e_\mathrm{s}\leq1$ \cite{Schaefer:1996}), but rather than being a constant, it depends on the collision velocity \cite{Brilliantov:1996,Schwager:2008} and angle \cite{Luding:1998}. A general, physically accurate expression for a damping coefficient $c_\mathrm{s}(e_\mathrm{s})$ with constant $e_\mathrm{s}$ is therefore impossible to formulate, and some DEM implementations just set $c_\mathrm{s}=c_\mathrm{n}$ \cite{Tsuji:1992,Zhou:1999,Silbert:2002,Remy:2009,Incardona:2019,Santos:2020,Qian:2022,Qian:2023} or use fixed ratios $c_\mathrm{s}/c_\mathrm{n}$ \cite{Silbert:2001,Ringl:2012,Garg:2012}. Here, a heuristic approach it taken instead: Eq.~\ref{eq:cs} is used with a factor $\sqrt{1/6}$ determined numerically in a direct central collision with pure sliding ($\vec{u}_\mathrm{s}=\vec{0}$), in such a way that $0\leq e_\mathrm{n}=e_\mathrm{s}\leq1$ over the entire range of values. With $\vec{u}_\mathrm{s}$ included, the tangential contact velocity can then reverse (corresponding to $e_\mathrm{s}<0$) in collisions with sufficient static friction, as it should be \cite{Luding:1998}. It appears that this expression for $c_\mathrm{s}$ has not occurred in the DEM literature before.

For simplicity, \program{} assumes that all particles have identical material properties ($E_i\equiv E$, $\nu_i\equiv\nu$, $\rho_i\equiv\rho$). Some of the above expressions therefore simplify:
\begin{align}
	E^* &= \frac{E}{2(1-\nu^2)},\\
	G^* &= \frac{E}{4(2-\nu)(1+\nu)},\\
	m^* &= \frac{4}{3}\pi\rho\left(\frac{1}{R_i^3}+\frac{1}{R_j^3}\right)^{-1},\\
	c_\mathrm{s} &= c_\mathrm{n}\sqrt{\frac{1-\nu}{10(2-\nu)}}.
\end{align}
Note that the ratio between tangential and normal damping $c_\mathrm{s}/c_\mathrm{n}$ is confined to the range $(0.18,0.26)$ for a collision between two equal particles here, for any Poisson ratio $\nu\in(-1,0.5]$.

\program{} handles particle contact with static walls and other geometric primitives (detailed further below), all by the same routine. For a collision between a particle $i$ and a static object $j$, the latter is assumed to be infinitely flat, stiff and heavy ($R_j=\infty$, $E_j=\infty$, $m_j=\infty$), such that $R^*=R_i$, $m^*=m_i$, $E^*=E_i/(1-\nu_i^2)$, and $G^*=G_i/(2-\nu_i)$.

To obtain the normal and tangential velocity at the contact point for Eqs.~\ref{eq:fn} and \ref{eq:fs}, the total relative velocity $\vec{v} = \vec{v}_\mathrm{n} + \vec{v}_\mathrm{s}$, is decomposed as
\begin{align}
	\vec{v}& = \vec{v}_i - \vec{v}_j + \vec{\hat{n}} \times (r_i\vec{\omega}_i - r_j\vec{\omega}_j), \\
	\vec{v}_\mathrm{n} &= \left(\vec{v} \cdot \vec{\hat{n}} \right) \vec{\hat{n}}, \\
	\vec{v}_\mathrm{s} &= \vec{v} - \vec{v}_\mathrm{n},
\end{align}
where $\vec{v}_{i}$, $\vec{\omega}_{i}$ and $\vec{v}_{j}$, $\vec{\omega}_{j}$ are the linear and angular velocities of bodies $i$ and $j$ at and about their centers of mass, respectively. Walls are assumed to be stationary ($\vec{v}_j=\vec{0}$, $r_j\vec{\omega}_j=\vec{0}$). The radii
\begin{align}
	r_i &= R_i-\delta/2,\\
	r_j &= R_j-\delta/2
\end{align}
measure the distance from the centers of mass of bodies $i$ and $j$ to the mutual contact point (Fig.~\ref{fig:3}).

The form of the torques in Eq.~\ref{eq:rotation} has been the subject of considerable debate in the literature \cite{Ai:2011,Wang:2015,Zhao:2016,Zhao:2018}. Rolling resistance is a complex subject with several different models that have been proposed \cite{Ai:2011,Wang:2015}. Of the three torques, the twisting resistance is often considered the least important, and therefore neglected in many codes \cite{Jiang:2015,Santos:2020}. However, both rolling and twisting friction can be required to reproduce experimentally generated granular packings \cite{Santos:2020}. In \program{}, the sliding, rolling and twisting torques are implemented as \cite{Luding:2008,Marshall:2009}
\begin{align}
	\vec{\tau}_\mathrm{s} &= r_i\vec{f}_\mathrm{s}\times\vec{\hat{n}},\\
	\vec{\tau}_\mathrm{r} &= R^*\vec{\hat{n}}\times\vec{f}_\mathrm{r},\\
	\vec{\tau}_\mathrm{t} &= f_\mathrm{t}R^*\vec{\hat{n}},
\end{align}
with linearly viscoelastic (Kelvin--Voigt) rolling and twisting forces:
\begin{align}
	\vec{f}_\mathrm{r} &= -k_\mathrm{r}\vec{u}_\mathrm{r}-c_\mathrm{r}\vec{v}_\mathrm{r},\label{eq:fr}\\
	f_\mathrm{t} &= -k_\mathrm{t}u_\mathrm{t}-c_\mathrm{t}v_\mathrm{t},\label{eq:ft}
\end{align}	
where $\vec{u}_\mathrm{r}$ and $u_\mathrm{t}$ are the integrated displacements of static rolling and twisting friction, respectively, defined later. The relative rolling and twisting velocities can be obtained as \cite{Luding:2008,Wang:2015}
\begin{align}
	\vec{v}_\mathrm{r} &= \left(\frac{1}{r_i}+\frac{1}{r_j}\right)^{-1}\vec{\omega}\times\vec{\hat{n}},\\
	v_\mathrm{t} &= R^*\vec{\omega}\cdot\vec{\hat{n}},\\
	\vec{\omega} &= \vec{\omega}_i-\vec{\omega}_j,
\end{align}
which also applies for particle-wall collisions ($r_j=\infty$).

It has been suggested that the twisting resistance parameters in Eqs.~\ref{eq:fr} and \ref{eq:ft} can be linked to those for sliding through integration of the frictional stress over the contact area \cite{Marshall:2009}. In the notation used here, these relationships would read $k_\mathrm{t} = bk_\mathrm{s}$, $c_\mathrm{t} = bc_\mathrm{s}$ where $b=\delta/(2R^*)$ is the relative indentation depth. Similarly, the coefficients for rolling resistance may be related to the normal contact parameters through a dimensionless shape parameter \cite{Jiang:2015} which, when interpreted as the particle deformation from indentation, result in $k_\mathrm{r} = bk_\mathrm{n}/2$, $c_\mathrm{r} = bc_\mathrm{n}/2$. These expressions lead to relatively small stiffness and damping parameters for rolling and twisting and in practical tests with \program{}, prolonged oscillations with unrealistically large amplitudes were observed with them, unless substantial damping is used. Alternatively, rolling resistance can be related to the visco-elastic properties of the particles, and the coefficient of restitution that results from a collision with a given impact velocity \cite{Brilliantov:1998}. Iwashita and Oda \cite{Iwashita:1998} noted that ``there is no rational reason, unfortunately, to choose a specific value as the rolling stiffness $k_\mathrm{r}$''. Other DEM codes use fixed ratios between the coefficients for sliding, rolling and/or twisting \cite{Luding:2008} or set them equal \cite{Iwashita:1998,Santos:2020}. \program{} follows that heuristic by setting $k_\mathrm{r} = k_\mathrm{n}$, $k_\mathrm{t} = k_\mathrm{s}$, $c_\mathrm{r} = c_\mathrm{n}$, $c_\mathrm{t} = c_\mathrm{s}$.

Analogously to the sliding friction force (Eq.~\ref{eq:Coulombs}), also the rolling and twisting friction forces (Eqs.~\ref{eq:fr} and \ref{eq:ft}) are clipped to satisfy Coulomb's law, each with their own friction coefficients:
\begin{align}
	\norm{\vec{f}_\mathrm{r}} &\leq \mu_\mathrm{r}\norm{\vec{f}_\mathrm{n}},\label{eq:Coulombr}\\
	\abs{f_\mathrm{t}} &\leq \mu_\mathrm{t}\norm{\vec{f}_\mathrm{n}}\label{eq:Coulombt}.
\end{align}
The Coulomb friction coefficients for sliding and rolling, $\mu_\mathrm{s}$ and $\mu_\mathrm{r}$, are independent material parameters, but the one for twisting can be related to sliding as $\mu_\mathrm{t}=2\mu_\mathrm{s}/3$ \cite{Marshall:2009}, because the twisting motion is essentially rotational sliding.

With these expressions, the inelastic contact model is completely defined and parameterized, with exception of the calculation of the relative displacements in shear, rolling and twisting resistance, $\vec{u}_\mathrm{s}$, $\vec{u}_\mathrm{r}$, $u_\mathrm{t}$ (i.e., the static friction history). These are initialized to zero at the beginning of a new collision and then updated in every timestep as detailed in the next section.

\subsection{Time integration}

To evolve the particle ensemble in time, the equations of motion are solved with the semi-implicit Euler method. Apart from its simplicity, it has the advantage of being symplectic, i.e., preserving the Hamiltonian when all dissipative forces are disabled ($e_\mathrm{n}=1$, $\mu_\mathrm{s}=\mu_\mathrm{r}=\mu_\mathrm{t}=\eta=0$). From the linear and angular accelerations of all particles,
\begin{align}
	\vec{a}_i &= \frac{1}{m_i}\sum_{j\neq i}\left(\vec{f}_\mathrm{n}^{ij}+\vec{f}_\mathrm{s}^{ij}\right) + \vec{g} - c_i\vec{v}_i,\\
	\vec{\alpha}_i &= \frac{1}{I_i}\sum_{j\neq i}\left(\vec{\tau}_\mathrm{s}^{ij} + \vec{\tau}_\mathrm{r}^{ij} + \vec{\tau}_\mathrm{t}^{ij}\right) - \gamma_i\vec{\omega}_i,
\end{align}
the velocities, positions and angular velocities are updated according to
\begin{align}	
	\vec{v}_i &\gets \vec{v}_i + \Delta t\,\vec{a}_i,\label{eq:v}\\
	\vec{x}_i &\gets \vec{x}_i + \Delta t\,\vec{v}_i,\label{eq:x}\\
	\vec{\omega}_i &\gets \vec{\omega}_i + \Delta t\,\vec{\alpha}_i.\label{eq:w}
\end{align}
where $\Delta t>0$ is the finite timestep. Note that the order in which Eqs.~\ref{eq:v} and \ref{eq:x} are evaluated matters: To be symplectic, the position updates (Eq.~\ref{eq:x}) must use the already updated velocities (Eq.~\ref{eq:v}).

\begin{table*}
\centering
\caption{\textbf{Complete list of program parameters.} In the parameter dimension, M represents mass, L length, T time. Default values create a quick simulation of a small polydisperse ensemble of soft particles falling under gravity through an hourglass (Fig.~\ref{fig:5}A).}
\label{tab:2}
\begin{tabular}{lrccl}
\toprule
Symbol & Default value & Requirements & Dimension & Description\\
\midrule
\multicolumn{5}{l}{\textbf{Geometric parameters}}\\
$R_\mathrm{min}$ & $0.002$ & $>0$ & L & Minimum particle radius\\
$R_\mathrm{max}$ & $0.01$ & $\geq R_\mathrm{min}$ & L & Maximum particle radius\\
$R_\mathrm{mesh}$ & $\{0.002, 0.02, 0.01\}$ & $\geq0$ & L & Mesh radii (can be a list of values)\\
$\vec{s}_\mathrm{min}$ & $(-0.1,-0.1,0.3)$ & --- & L & Minimum coordinate of particle spawn box\\
$\vec{s}_\mathrm{max}$ & $(0.1,0.1,0.4)$ & $\geq\vec{s}_\mathrm{min}$ & L & Maximum coordinate of particle spawn box\\
\\
\multicolumn{5}{l}{\textbf{Material parameters}}\\
$\rho$ & $10^3$ & $>0$ & M/L\textsuperscript{3} & Mass density\\
$E$ & $10^6$ & $>0$ & M/LT\textsuperscript{2} & Young's modulus\\
$\nu$ & $0.3$ & $\in(-1,0.5]$ & --- & Poisson's ratio\\
$e_\mathrm{n}$ & $0.5$ & $\in[0,1]$ & --- & Coefficient of normal restitution\\
$\mu_\mathrm{s}$ & $0.3$ & $\geq0$ & --- & Coulomb friction coefficient for sliding\\
$\mu_\mathrm{r}$ & $0.3$ & $\geq0$ & --- & Coulomb friction coefficient for rolling\\
$\mu_\mathrm{t}$ & $2\mu_\mathrm{s}/3$ & $\geq0$ & --- & Coulomb friction coefficient for twisting\\
$\eta$ & $0$ & $\geq0$ & M/LT & Dynamic viscosity for Stokes drag\\
\\
\multicolumn{5}{l}{\textbf{Kinetic parameters}}\\
$\vec{v}_0$ & $(0,0,-1)$ & --- & L/T & Initial linear particle velocity\\
$\vec{w}_0$ & $(0,0,0)$ & --- & 1/T & Initial angular particle velocity\\
$\vec{g}$ & $(0,0,-9.81)$ & --- & L/T\textsuperscript{2} & Gravitational acceleration\\
\\
\multicolumn{5}{l}{\textbf{Simulation parameters}}\\
$\Delta t$ & Eq.~\ref{eq:dt} & $>0$ & T & Time increment\\
$N_\mathrm{max}$ & $1000$ & $\geq0$ & --- & Maximum number of particles\\
$N_\mathrm{frames}$ & $100$ & $\geq0$ & --- & Number of output frames\\
$N_\mathrm{steps}$ & $500$ & $\geq0$ & --- & Number of timesteps per frame\\
$N_\mathrm{spawn}$ & $100$ & $\geq0$ & --- & Number of particle spawn attempts per timestep\\
\bottomrule
\end{tabular}
\end{table*}

The actual orientation of spherical particles does not necessarily need to be tracked, as all that is needed to evaluate the forces are angular velocities, not angles. Nevertheless, for various purposes including visualization (or to generalize the code to non-spherical particles), it can be useful to store and update particle orientations, too. Updating them using Euler angles would lead to the infamous gimbal lock, the inability to rotate in certain directions when two rotational axes coincide. A simple and robust way to avoid this problem is to use unit quaternions to represent particle orientations, although they are not the only possible remedy \cite{Campello:2015}. A quaternion can be written as $\hat{q}=(q,\vec{q})$ where $q$ is the scalar part and $\vec{q}$ the vectorial part. Numerous algorithms exist to integrate quaternions in time. \program{} uses the synchronous version of a recently introduced, highly accurate method called SPIRAL \cite{delValle:2024}:
\begin{equation}
\label{eq:spiral}
	\hat{q}_i \gets \hat{q}_i\left(\cos\varphi_i,\,\sin\varphi_i\frac{\vec{\omega}_i}{\norm{\vec{\omega}_i}}\right)\left(\cos\theta_i,\,\sin\theta_i\frac{\vec{\alpha}_i}{\norm{\vec{\alpha}_i}}\right)
\end{equation}
where
\begin{equation}
	\varphi_i = \frac{\Delta t}{2}\norm{\vec{\omega}_i}, \qquad \theta_i = \left(\frac{\Delta t}{2}\right)^2\norm{\vec{\alpha}_i}.
\end{equation}
The triple product in Eq.~\ref{eq:spiral} is computed from left to right as two consecutive Hamilton products:
\begin{equation}
	\hat{p}\hat{q} = \left(pq-\vec{p}\cdot\vec{q},\,p\vec{q}+q\vec{p}+\vec{p}\times\vec{q}\right).
\end{equation}
Note that these products are computed only if $\norm{\vec{\omega}_i}>0$ and $\norm{\vec{\alpha}_i}>0$, respectively. Moreover, note that the quaternion must be updated before the angular velocity, i.e., before Eq.~\ref{eq:w} is evaluated.\\

For reasonable numerical stability and accuracy with explicit single-step time integration, the size of the timestep usually needs be chosen well below the period of the fastest harmonic oscillator in the Newtonian system: $\Delta t\ll\sqrt{m/k}$, where $m$ is a typical (small) mass and $k$ a typical (large) stiffness. A good approach is to base the size of the timestep on the propagation of elastic waves across particles, which need to be resolved with sufficient resolution \cite{Li:2005, Burns:2019}. In \program{}, a value of
\begin{equation}
\label{eq:dt}
\Delta t=R_\mathrm{min}\sqrt{\rho\frac{1-\nu^2}{E}}
\end{equation}
is used by default, resolving a head-on normal collision of two equal particles with a maximum relative indentation depth $\delta/(2R^*)=10\%$ with a dozen timesteps at $e_\mathrm{n}=1/2$. $R_\mathrm{min}$ is the specified minimum particle size in the (possibly polydisperse) ensemble. For particles with a radius of 1\,cm, a mass density of 1\,g/cm\textsuperscript{3}, and a Young's modulus of 1\,MPa, the timestep will be about 0.3\,ms, and the simulation of a minute physical time thus requires about 200,000 timesteps. Smaller $\Delta t$ values may be set by the user if higher accuracy is desired, for example to accurately resolve stiff forces that can result from large friction coefficients.\\

In the evaluation of the frictional forces (Eqs.~\ref{eq:fs}, \ref{eq:fr} and \ref{eq:ft}), the displacement of the static friction springs,
\begin{equation}
\label{eq:um}
\vec{u}_\mathrm{m}(t) = \int_{t_0}^t \vec{v}_\mathrm{m}(t')\,dt',
\end{equation}
(integrated in the co-rotated frame since the time of first contact $t_0$), needs to be updated for each contact, for each friction mode $\mathrm{m}\in\{\mathrm{s},\mathrm{r},\mathrm{t}\}$ (sliding, rolling and twisting). From the previous timestep, the sliding and rolling displacements are first projected back onto the tangent plane perpendicular to the current unit normal vector $\vec{\hat{n}}$, and scaled back to their original length unless zero \cite{Luding:2008}:
\begin{equation}
\vec{u}_\mathrm{m} \gets \norm{\vec{u}_\mathrm{m}}\frac{\vec{u}_\mathrm{m} - \left(\vec{u}_\mathrm{m} \cdot \vec{\hat{n}} \right) \vec{\hat{n}}}{\norm{\vec{u}_\mathrm{m} - \left(\vec{u}_\mathrm{m} \cdot \vec{\hat{n}} \right) \vec{\hat{n}}}},\quad\mathrm{m}=\mathrm{s},\mathrm{r}.
\end{equation}
Note that for the twisting friction, a projection onto the current normal is not needed, as it is readily expressed as a scalar torque about $\vec{\hat{n}}$. The frictional forces are then evaluated using these updated spring displacements. If the Coulomb inequality (Eqs.~\ref{eq:Coulombs}, \ref{eq:Coulombr} and \ref{eq:Coulombt}) is satisfied, friction is in the static regime, and the displacement is incremented for the next timestep as
\begin{equation}
\label{eq:um_static}
\vec{u}_\mathrm{m} \gets \vec{u}_\mathrm{m} + \Delta t\,\vec{v}_\mathrm{m},\quad\mathrm{m}=\mathrm{s},\mathrm{r},\mathrm{t}.
\end{equation}
Otherwise, i.e., in the dynamic friction regime, the spring is set to the length that fulfills Coulomb's condition equally \cite{Luding:2008}:
\begin{align}
\vec{f}_\mathrm{m} &\gets \mu_\mathrm{m}\norm{\vec{f}_\mathrm{n}}\frac{\vec{f}_\mathrm{m}}{\norm{\vec{f}_\mathrm{m}}},\label{eq:fm_kinetic}\\
\vec{u}_\mathrm{m} &\gets -\frac{1}{k_\mathrm{m}}\left(\vec{f}_\mathrm{m} + c_\mathrm{m} \vec{v}_\mathrm{m}\right),\quad\mathrm{m}=\mathrm{s},\mathrm{r},\mathrm{t},\label{eq:um_kinetic}
\end{align}
forgetting about any potential previous history of $\vec{u}_\mathrm{m}$. This ensures a continuous force transition across the switch from dynamic to static friction. Note that Eqs.~\ref{eq:um}, \ref{eq:um_static}, \ref{eq:fm_kinetic} and \ref{eq:um_kinetic} are to be interpreted in scalar form for twisting ($u_\mathrm{t}$, $v_\mathrm{t}$, $f_\mathrm{t}$ in place of $\vec{u}_\mathrm{t}$, $\vec{v}_\mathrm{t}$, $\vec{f}_\mathrm{t}$). The procedure is applied for each of the three friction modes m individually, not together, such that they can be in static or kinetic state independently from each other. The static and kinetic friction modes are assumed to have the same friction coefficients here---a simplification commonly made in DEM codes that implement Coulomb friction, such as \cite{Luding:2008,Kloss:2011,Ringl:2012,Santos:2020}.

\section{Implementation}

\program{} was designed following a min-max coding philosophy, providing as much typical core DEM functionality (for a certain set of common use cases) with as little and as comprehensible code as possible, with goals like leanness, ease of maintenance, and portability in mind. One example manifestation of this is the floating-point precision used in all calculations, which can be changed with a single \texttt{typedef}. The preset precision is \texttt{double}. The entire program consists of only two C++ files, \texttt{tinydem.hpp} and \texttt{tinydem.cpp}, totaling in about 600 lines of commented code. For maximum accessibility, variables are largely named as written in this paper. The header file contains only generic and minimal data structures, distance calculation and I/O routines, and normally need not be modified by the user. The source file contains the physical model with all its parameters (Table~\ref{tab:2}), the simulation loop, etc., which may be altered by the user to simulate a specific scenario.

\subsection{Particle generation}

\program{} supports two ways of generating particles: Through an initial import from a CSV file (see usage instructions below), or through dynamically spawning of new particles within a defined spatial region. To use the second method, a spawn box can be defined via its lower and upper corner vertices, $\vec{s}_\mathrm{min}$ and $\vec{s}_\mathrm{max}$. Additionally, a number of particle spawn attempts per timestep, $N_\mathrm{spawn}$, can be specified. In each attempt, a random particle center position is drawn uniformly within the spawn box, a random particle radius is drawn uniformly in the range $[R_\mathrm{min},R_\mathrm{max}]$, and unless this would result in an overlap with other particles or the mesh, the particle is generated with initial velocity $\vec{v}_0$ and angular velocity $\vec{\omega}_0$ there. Note that in case of failure to spawn the particle at the drawn location, the assigned random radius is carried over to the next particle spawn attempt (possibly during the next timestep) to avoid biasing the particle size distribution toward smaller particles that have a higher chance of filling small gaps. Particles are spawned over time only until a specified maximum number $N_\mathrm{max}$ is reached.

\program{} can be run in pure 2D mode in any plane (or multiple planes) parallel to any of the three Cartesian planes, or in 1D mode along any line (or multiple lines) parallel to the Cartesian axes. As long as all forces, torques and velocities are restricted to such a plane or line, particles will remain confined to it. All that is needed for this is that the initial conditions (including the particle import file and particle spawn region) are conforming, and that the mesh contains no elements positioned such that they can collide with the particles in an oblique direction, pushing them out of their plane or line.

Note that, when generating particles with the spawn region method, only the particles' center points, not their entire volumes, are restricted to lie within the specified spawn region. This is to enable lower-dimensional (2D or 1D) spawn regions also for polydisperse ensembles by setting a subset of the $x$-, $y$- or $z$-components of the spawn box vertices $\vec{s}_\mathrm{min}$ and $\vec{s}_\mathrm{max}$ equal.

\subsection{Particle and collision data}

\begin{table}
\centering
\caption{\textbf{Summary of particle-specific variables implemented.} Particle indices are omitted for readability. Particle properties not listed (such as elastic moduli and mass densities) are shared among all particles and thus do need not be stored per particle. In the dimension, L represents length, T time.}
\label{tab:3}
\begin{tabular}{@{}ll@{}cl@{}}
\toprule
Symbol & Type & Dimension & Description\\
\midrule
\multicolumn{4}{@{}l}{\textbf{Particle-specific variables}}\\
$R$ & scalar & L & radius\\
$\vec{x}$ & vector & L & position\\
$\vec{v}$ & vector & L/T & velocity\\
$\vec{a}$ & vector & L/T\textsuperscript{2} & acceleration\\
$q$ & scalar & --- & scalar part of quaternion\\
$\vec{q}$ & vector & --- & vectorial part of quaternion\\
$\vec{\omega}$ & vector & 1/T & angular velocity\\
$\vec{\alpha}$ & vector & 1/T\textsuperscript{2} & angular acceleration\\
$n$ & integer & --- & index of next particle in cell\\
$c$ & list & --- & list of contacts (entries below)\\
\\
\multicolumn{4}{@{}l}{\textbf{Contact-specific variables} (per entry of $c$ above)}\\
$j$ & integer & --- & index of contact partner\\
$\vec{u}_\mathrm{s}$ & vector & L & sliding spring displacement\\
$\vec{u}_\mathrm{r}$ & vector & L & rolling spring displacement\\
$u_\mathrm{t}$ & scalar & L & twisting spring displacement\\
$\psi$ & Boolean & --- & flag marking active collisions\\
\bottomrule
\end{tabular}
\end{table}

For a better overview, Table~\ref{tab:3} lists all properties stored per particle object. Note that the contact forces and torques are antisymmetric,
\begin{align}
	\vec{f}_\mathrm{n}^{ij} &= -\vec{f}^{ji}_\mathrm{n},\label{eq:antisymm_normal}\\
	\vec{f}_\mathrm{s}^{ij} &= -\vec{f}^{ji}_\mathrm{s},\\
	\vec{\tau}_\mathrm{r}^{ij} &= -\vec{\tau}^{ji}_\mathrm{r},\\
	\vec{\tau}_\mathrm{t}^{ij} &= -\vec{\tau}^{ji}_\mathrm{t},
\end{align}
except the sliding torque, for which the relationship
\begin{equation}
\label{eq:antisymm_sliding}
	r_j\vec{\tau}_\mathrm{s}^{ij} = r_i\vec{\tau}^{ji}_\mathrm{s}
\end{equation}
holds \cite{Luding:2008}. One may therefore iterate over the ordered body pairs $i<j$ only once, computing the forces and torques on both of them using these relationships from a single interaction, or over the unordered pairs $i\neq j$ twice, reevaluating them from both of the particles' perspectives. In \program{}, the single-evaluation approach is taken, exploiting Eqs.~\ref{eq:antisymm_normal}--\ref{eq:antisymm_sliding}, as it is more efficient.

Each particle $i$ holds a list $c$ of contact data structures for collisions with particles $j$ with $i<j$. That way, since $j>0$ for particle-particle collisions by this principle, non-positive indices $j\leq0$ are free to be used to label collisions with mesh elements. Each entry in $c$ of particle $i$ holds the index $j$ of the contact partner, the static friction spring displacements, and a Boolean flag $\psi\in\{0,1\}$ indicating whether the collision is still active in the current timestep. During the dense phase of collision detection (detailed below), all active collisions are marked ($\psi=1$). Once all collisions involving a particle $i$ have been handled, all inactive contact objects ($\psi=0$) of that particle that are still present from the previous timestep are destroyed and the flag of all remaining (i.e., active) ones is reset for the next iteration ($\psi=0$).

\subsection{Collision detection}

\begin{figure*}
	\centering
	\includegraphics[width=0.85\linewidth]{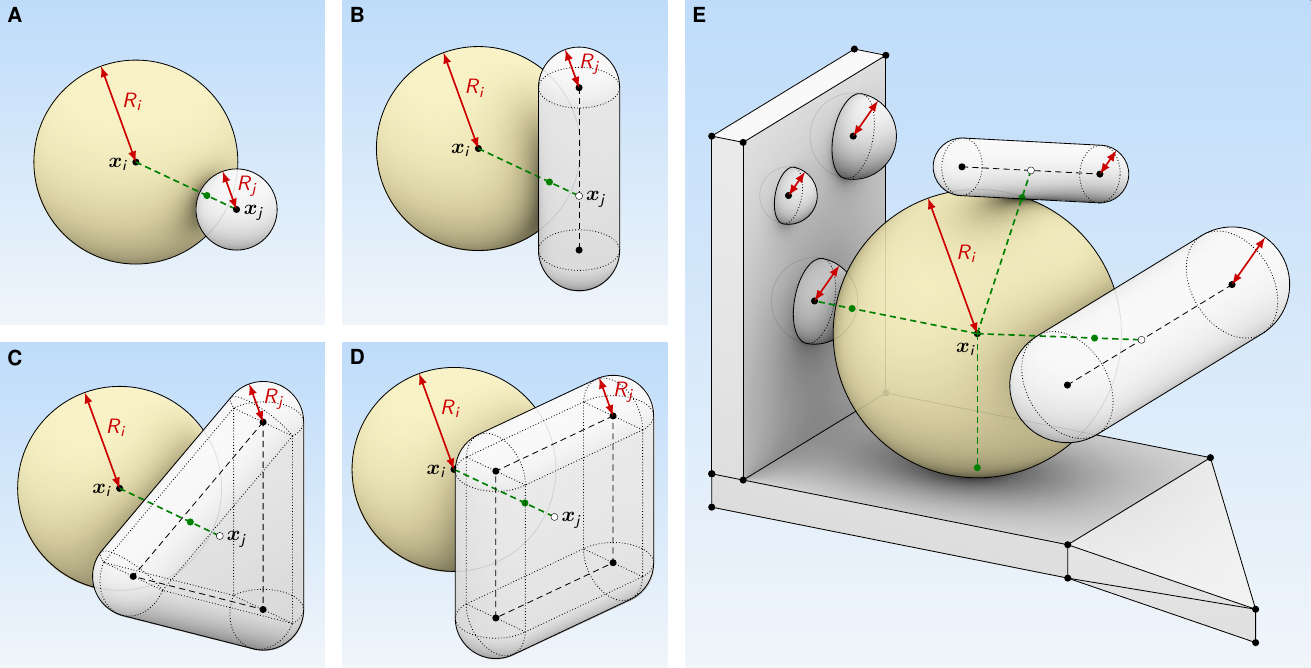}
	\caption{\textbf{Particle-mesh collisions.} Collisions between particles (yellow) and four different mesh primitives (white) are implemented: a point (A), line segment (B), triangle (C), and rectangle (D). With an optional mesh ``radius'' $R_j\geq0$ (red), each mesh element $j$ can be given an individual thickness with rounded-off edges. The contact point is shown in green, the closes point of approach on the mesh (which may also lie on the dashed edges of the mesh primitive) in white. By combining these elements, complex mesh geometries can be formed (E), allowing for concurrent contacts of a particle with multiple mesh elements, each having their own ``radius''. Mesh vertices are shown as black dots. The triangular and rectangular mesh elements have $R_j=0$ in E, hence the sharp edges.
	}
	\label{fig:4}
\end{figure*}

To handle particle collisions efficiently with linear time complexity in the number of particles $N$, it is essential to avoid unnecessary long-distance collision checks. \program{} uses regular space partitioning with linked cell lists \cite{Quentrec:1973}. In each timestep, the bounding box of the entire particle ensemble is calculated and discretized into cubic cells with an edge length equal to the specified maximum particle diameter, $2R_\mathrm{max}$. This ensures that all potential contact candidates can be found within the local Moore neighborhood of up to 27 cells about each particle. Only particle pairs within this neighborhood are then tested for overlap, jumping from one particle to the next one in the same cell (Table~\ref{tab:3}).

Note that the memory used by the spatial grid scales with its volume. Binary space partitioning with a tree data structure would prevent this, but is a complexity eschewed here for simplicity. Therefore, it is generally advisable to spatially confine the ensemble by defining a global bounding box with the mesh functionality detailed below to prevent excessive memory usage in simulations where particles would otherwise vastly separate.

In addition to collisions among particles, \program{} supports contact with arbitrarily shaped static meshes, such as walls or other obstacles. They are unlimited in number, may be connected or disjoint, and can consist of polygonal elements with one to four vertices. Thus, any discrete object whose surface consists of points (Fig.~\ref{fig:4}A), edges (Fig.~\ref{fig:4}B), triangles (Fig.~\ref{fig:4}C) and rectangles (Fig.~\ref{fig:4}D) can be used to define a geometrical environment (Fig.~\ref{fig:4}E) for the particle simulation. Like the particles, each mesh element $j$ can have its own radius $R_j\geq0$, and their edges and vertices are rounded off accordingly, i.e., a mesh vertex is effectively treated as a sphere, an edge as a spherocylinder, etc. The geometrical part of the particle-mesh collision detection thus reduces to four types of primitive subproblems: computing the distance vector between a point and 1) another point; 2) a line segment; 3) a triangle; and 4) a rectangle (Fig.~\ref{fig:4}). For all of these, simple direct algorithms are commonly available \cite{Ericson:2005} and implemented in \program{}.

A rather efficient way to implement the broad phase of collision detection between particles and mesh elements is to precompute each mesh element's axis-aligned bounding box, and then dynamically determine the indices of the block of grid cells it overlaps with. The dense phase of actual distance calculations can then be limited to the (usually small fraction of) particles that lie in those cells $\pm1$. Mesh elements that are far from any particle are thus not entering the dense phase at all. This leads to a time complexity for collisions between the $N$ particles and $M$ mesh elements that is linear in $M$ and sub-linear in $N$, unless the mesh densely fills the space occupied by the particle ensemble.

\begin{figure*}
	\centering
	\includegraphics[width=\linewidth]{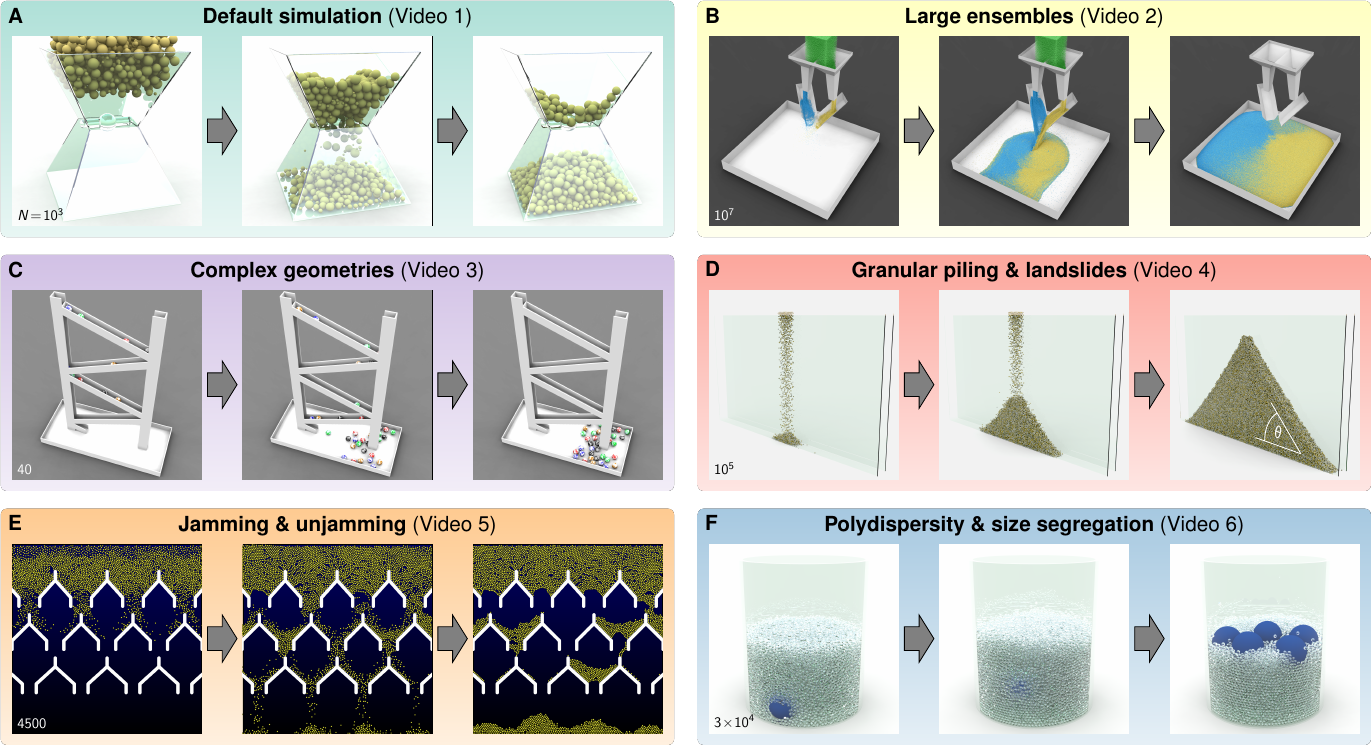}
	\caption{\textbf{Example simulations.} Particle numbers $N$ are given in the lower left corners. All six simulations have accompanying supplementary videos as labeled.
	A: The default simulation as preconfigured in the supplied code, a polydisperse set of particles falling through an hourglass with additional obstacles at the throat.
	B: Binary mixture of two streams consisting of 10 million grains.
	C: Marble run to demonstrate particle rotations and a somewhat more complex mesh geometry.
	D: A quasi-2D pile of strongly frictional grains between two vertical plates. $\theta$ is the angle of repose.
	E: Jamming in 2D. Frictional particles fall through three layers of funnels with neck sizes of 10, 8 and 6 particle diameters, from top to bottom.
	F: Particle size segregation through vibration of the base of a cylindrical container. Seven balls (blue, initially placed on the ground) have a ten-fold diameter than the rest of the beads, but equal material properties. Images show the gradual upward convection of the larger balls after 0, 83 and 175 vibration cycles.
	}
	\label{fig:5}
\end{figure*}

A subtle difficulty arises when a particle moves from a collision with one mesh element to another, for instance across the internal shared edge between two coplanar triangles making up a planar wall. For a period proportional to $R_j/\norm{\vec{v}_i}$, the particle will experience a repulsive force from both elements, unless dedicated countermeasures are taken. Such countermeasures are a non-trivial task, though: It could well be that the particle is righteously in contact with multiple mesh elements at the same time (for example in the corner of a box, or rolling down an inclined furrow)---a case that would need to be distinguished from rolling or sliding across seams of a flat mesh surface, which is a single continuous collision event. This problem of ``ghost collisions'' is well-known in rigid-body simulations and has no known universal, computationally efficient solution \cite{Havok:2005}. To mitigate the error made in the contact force \emph{magnitude}, weighting based on intersection areas can be used \cite{Fleissner:2007}, but this does not ensure a correct force \emph{direction}. A robust method based on Voronoi regions has been proposed \cite{Terdiman:2015}, but requires multiple passes, mesh connectivity information, and a distinction between sphere-polygon contact locations (face, edge or vertex)---a degree of complexity intentionally avoided in \program{}. Instead, a simple approach is implemented: particles can collide with any number of mesh elements simultaneously. This avoids the complexity and cost of continuous collision detection algorithms at the expense of losing contact information (Table~\ref{tab:3}) when a contact crosses over from one mesh element to another. Avoiding the ghost collisions across element borders as much as possible is one of the reasons why rectangular mesh elements are implemented here, even though they could also be represented by two joined triangles. The user is advised that ghost bumps can occur in the provided implementation, and that meshes should be as coarse as possible to reduce them to a minimum.

\section{Example simulations}

To demonstrate the range of potential applications of \program{}, six classical simulations are briefly showcased in Fig.~\ref{fig:5}.

The first scenario presents the default parameter setting used when \program{} is run out of the box. A small polydisperse ensemble of 1000 soft particles is simulated, spawning and dropping under the effect of gravity into an houglass-like container with a cylindrical and a spherical obstacle partially blocking the throat (Fig.~\ref{fig:5}A, Video~1). In the simulated period of about three physical seconds, the particles fall through to the bottom or get jammed, depending on randomness in particle initialization and in the race of threads in parallel execution. No viscous drag is used; the particles come to rest only through inelastic and frictional collisions with each other and with the mesh. The complete list of simulation parameters is given in Table~\ref{tab:2}. The simulation takes about half a minute on a 2.3 GHz Intel Core i9 processor with 8 threads.

The purpose of the second example is to demonstrate the suitability of \program{} to simulate phenomena involving large numbers of particles, despite its simple implementation. A stream of equal grains is dropped into a separator funnel that dyes them with two different colors (Fig.~\ref{fig:5}B, Video~2). Additional grains are dynamically added to the stream as room becomes available inside the spawn region at the top. The two sub-streams then fall onto chutes that merge them again, resulting in a binary dispersion pattern in the particle bed below. Collisions are frictional and inelastic, but no viscous drag is used. The simulation was run on an 8-core CPU and terminated when $N_\mathrm{max}=10$~million particles were reached.

As a third scenario, a marble run is simulated to showcase a somewhat more complex geometric obstacle course. \program{} does not pose any limits to the complexity of the static mesh, and is therefore suited also for more complicated geometric setups than shown here. With 40 texturized marbles completing the run, this example also shows the quaternion-based particle rotation in action (Fig.~\ref{fig:5}C, Video~3). The marbles are modeled without rolling friction and increased coefficient of restitution, but with moderate viscous drag.

The fourth simulation demonstrates a classical test of static friction in DEM codes. Confined by two parallel vertical plates separated by 11 grain diameters, a stream of 100,000 monodisperse grains piles up into a triangular heap (Fig.~\ref{fig:5}D, Video~4). Due to static friction, both sides form a stable linear slope (up to a logarithmic deviation at the bottom ends, which is a boundary effect \cite{Alonso:1996}). Greater angles produce landslides that flatten the slope, smaller angles let additional grains stack up to steepen it. Large friction coefficients ($\mu_\mathrm{s}=\mu_\mathrm{r}=3$) are used to produce a challenging simulation with steep fronts. The angle of repose $\theta$ yields an effective Coulomb friction coefficient $\mu_\mathrm{eff}=\tan\theta$ that is considerably smaller than $\mu_\mathrm{s}$, because the grains can rotate. Here, $\mu_\mathrm{eff}\approx1.15$ ($\theta\approx49^\circ$) is observed. Note that also the wall separation affects the slope \cite{Grasselli:1997, Zhou:2001}. To accurately resolve the stiff static friction forces, a tenfold lower timestep is used than set by default via Eq.~\ref{eq:dt}.

A classical use case of the DEM is the study of jamming and unjamming \cite{Behringer:2019}---a good opportunity to showcase \program{}'s 2D simulation mode. Although particles also jam in the default 3D setup (Fig.~\ref{fig:5}A), the behavior is better observable in 2D. About 4500 equally sized particles are spawned above three layers of lined-up funnels with narrowing throats. Under their own weight, particles spontaneously bridge the orifices (Fig.~\ref{fig:5}E, Video~5). This form of arching requires the throats to be no more than about 5 particle diameters wide at moderate friction \cite{Zuriguel:2005}, but stronger friction widens that critical size \cite{Pournin:2007}. The shown simulation does not use viscous drag but considerable friction ($\mu_\mathrm{s}=\mu_\mathrm{r}=1$) to promote jamming even with throat diameters of 6--10 particles. Although only anecdotal here, the formed arches are seen to have an aspect ratio (height to semi-width) of about one on average, consistent with previous reports \cite{Garcimartin:2010}. Similar to the previous piling example, a 5-fold reduced timestep is used for finer resolution of the static friction dynamics.

Finally, a showcase of another extensively studied phenomenon in granular media: size-induced particle segregation (Fig.~\ref{fig:5}F, Video~6). 30,000 beads are placed in a cylindrical container with an inner radius of 39 bead radii $R_\mathrm{min}$. Seven of them are ten times larger ($R_\mathrm{max}=10R_\mathrm{min}$, blue balls) but have the same material properties (mass density etc.), and are randomly placed at the very bottom, buried by the smaller beads. Then, the base is harmonically vibrated in vertical direction with amplitude $A=R_\mathrm{min}$ and frequency $f$ such that the peak acceleration $A(2\pi f)^2$ is equal to $3g$, a commonly used value that promotes quick segregation \cite{Zhao:2019}. In this agitated bath, the larger balls are convected upward to the top over time, against one's intuition that the heavier particles should settle at the bottom. This size segregation is facilitated by particle-wall friction, while particle-particle friction has no substantial effect \cite{Elperin:1997, Sun:2006}. In the simulation shown here, $\mu_\mathrm{s}=\mu_\mathrm{r}=0.3$ is used for both types of contact.

\section{Computational performance and resource usage}

\begin{figure}
	\centering
	\includegraphics[width=0.9\linewidth]{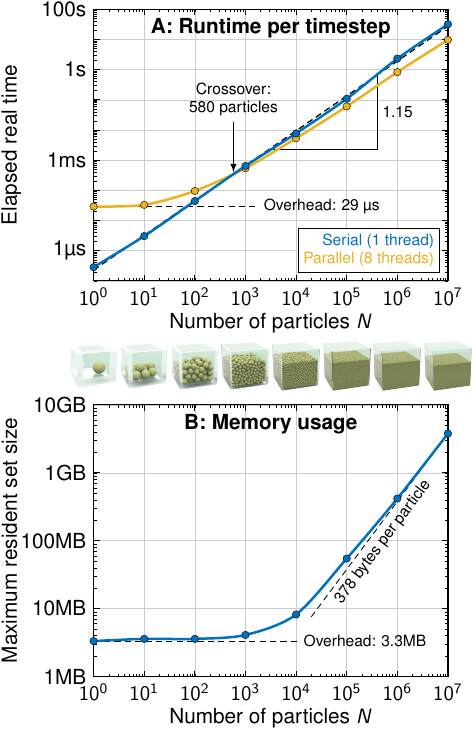}
	\caption{\textbf{Computational performance.}
	A: Scaling of the runtime per timestep with the number of particles. The dashed black slope is a fitted power law with annotated exponent.
	B: Scaling of the maximum memory used with the number of particles.
	Benchmarks were performed under Red Hat Enterprise Linux 9.4 on a 2019 Intel Xeon Gold 6244 CPU (3.60\,GHz, 8 cores). \program{} was compiled using GCC 11.2.0 with optimization level 3.
	}
	\label{fig:6}
\end{figure}

\program{} is not designed to max out high-performance computing infrastructure, but offers competitive performance on multi-core CPUs nevertheless. All loops over particle and mesh elements in the timestepping function are parallelized with OpenMP, with exception of the construction of the linked cell list, due to speedup limitations \cite{Halver:2016}. 

To measure the computational performance, a controlled and easily reproducible setting involving many particle contacts was chosen. $N$ particles were placed in a cubic container whose edge length scales as $N^{1/3}$, and allowed to settle under gravity. In this densely packed system, the wallclock time to perform one timestep was measured (averaged over at least a minute to buffer fluctuations) in serial mode and using 8 threads, for up to 10 million particles. At $N=1000$ particles, one timestep took about $0.64$\,ms on a 3.6 GHz Intel Xeon Gold 6244 CPU, which is more than an order of magnitude faster than the fastest Fortran implementation tested in 2006 on an 2.4 GHz AMD Athlon CPU in a similar benchmark \cite{Balevicius:2006}. This might just about reflect the performance differences between these CPUs. Although all code components scale at most linearly with $N$, slightly superlinear serial time complexity is observed in practice (Fig.~\ref{fig:6}A), presumably due to caching and memory bandwidth limitations: $\text{runtime}\sim N^\sigma$ with $\sigma=1.154\pm0.012$ (S.E.). Similar superlinear scaling (although with larger exponent $\sigma=1.33$) was previously found for comparable memory-bound vertex simulations \cite{Vetter:2024, Runser:2024}. In parallel execution, a crossover from parallelization overhead to speedup was observed at about 580 particles (interpolated). The number of particle updates per wall clock second---a quantity that became known as the Cundall number \cite{Hopkins:2004}---ranges from $3.5\!\times\!10^6$ to $3.2\!\times\!10^5$ in serial execution in this measurement.

In addition to the runtime, in can be useful to have an estimate of the expected memory requirements of a simulation. Fig.~\ref{fig:6}B shows the maximum memory occupied by \program{} in the same set of simulations. Since scenarios with fewer particle contacts require less memory (Table~\ref{tab:3}), the memory used in this high-density benchmark can be considered an upper bound for other reasonable scenarios, unless the ensemble is spread out enough to let the spatial partitioning grid eat up considerably more memory. The used memory was observed to transition from a constant overhead to linear scaling between about $N=10^3$ to $10^5$ particles. In the largest tested simulation, 378 bytes were used per particle. An ensemble of a million particles requires about 400 MB to simulate.
 
\section{Usage instructions}

In the spirit of \cite{Vetter:2024}, \program{} was developed to be highly portable and only requires a C++11 compiler. For multithreading support, the OpenMP 3.1 specification must be supported, which is the case in ICC since v12.1 (2011), in GCC since v4.7 (2012), and in Clang/LLVM since v3.7 (2015). To compile it, run
\begin{verbatim}
g++ -fopenmp -O3 -o tinydem tinydem.cpp
\end{verbatim}
or an equivalent command. Omitting the \texttt{-fopenmp} option compiles \program{} for serial execution. To run a simulation, set the desired parameters in lines 8--31 of \texttt{tinydem.cpp}, compile it, and execute the binary by typing
\begin{verbatim}
OMP_NUM_THREADS=8 ./tinydem mesh.off input.csv output
\end{verbatim}
or similar. If no number of threads is specified, the selection of a suitable number is delegated to OpenMP.

The three command-line arguments are optional. To use later arguments but skip previous ones, empty strings (\texttt{""}) can be used. With the first argument, the path to a text file in GeomView's Object File Format (OFF) \cite{OFF} can be given to define the mesh geometry. At the end of each face specification in the OFF file, an optional integer index can be specified. While this is intended as a colormap index in the OFF specification, \program{} uses it to specify the index in the list of mesh radii $R_\mathrm{mesh}$ (see the supplied default \texttt{mesh.off} file). If omitted, this index is set to 0, such that the first entry in $R_\mathrm{mesh}$ is the default mesh radius. The second argument is the path to a comma-separated value (CSV) file specifying the particles to start the simulation with. The format is identical to the output files generated by the program itself. For an example, run the default simulation. If none is given, the simulation starts with no particles, but still generates them with the spawning method described earlier. The third argument specifies an output directory. If unspecified, a folder named \texttt{output} is generated in the current working directory.

\program{} produces a time series of minimal CSV files for maximum compatibility. Existing output is overwritten without prompt. The number of files written is $1+N_\mathrm{frames}$, where $N_\mathrm{frames}$ is user-specified (Table~\ref{tab:2}) and the extra 1 is the initial state. Between frames, $N_\mathrm{steps}$ timesteps of size $\Delta t$ are performed, such that the total simulated physical time is given by $N_\mathrm{frames}N_\mathrm{steps}\Delta t$.

The simulated particle ensemble can be visualized in ParaView, for example. Open the CSV file series and apply the TableToPoints filter to it, using the $x$, $y$ and $z$ columns, with the ``Keep All Data Arrays'' option enabled. The particles can then be rendered either using the fast ``Point Gaussian'' representation with the radius $R$ selected in ``Use Scalar Array'', or using the 3D Glyphs representation with Sphere as the ``Glyph Type'', Magnitude as ``Scale Mode'' and $R$ as the ``Scale Array''. To also apply the quaternion rotation to the rendered particles (for instance to show their orientation with an Arrow glyph), apply a Programmable Filter after the TableToPoints filter with the following Python code:
\begin{minted}{python}
import numpy as np
q = [inputs[0].PointData["q%s"%i] for i in range(4)]
output.PointData.append(np.stack(q, axis=1), "q")
\end{minted}
This makes the quaternion available as ``Orientation Vectors'' in ParaView's 3D Glyphs.

\section{Conclusion}

Since about one and a half decades, open-source DEM codes have become almost a commodity. With a median number of lines of code beyond 90,000, however, the more than 20 available programs (Table~\ref{tab:1}) are complex and heavy, making it difficult to adapt them to one's own particular needs. Is writing capable custom DEM code a major undertaking, then, that the interested researcher or student should shy away from? Is it necessary to fine-tune numerous model parameters to obtain consistent physical behavior? It is not, as \program{} demonstrates.

\program{} is a lightweight standalone program to simulate frictional granular media that stands out of the crowd of existing DEM codes by combining two uncommon features:
\begin{itemize}
\item \textit{Accessibility.} \program{} consists of only just above 600 lines of compact, simple, commented C++ code in two files---a source and a header file. It is easy to handle, maintain, adjust or extend. Without dependencies, it is exceptionally independent and portable.
\item \textit{Comprehensive constitutive parameterization.} All three modes of frictional inelastic contacts (sliding, rolling and twisting) are implemented, and the latest constitutive relationships are used to simplify the parameterization to an intuitive set of fundamental material properties (Table~\ref{tab:2}).
\end{itemize}

\program{} is the only publicly available, full-fledged DEM program below 1,000 lines of code, and the only one below 10,000 lines that includes all three modes of torque exchange. All algorithms are direct and there are no numerical tolerances, magic numbers or similar. With its lean standalone implementation, maximal portability, and permissive 3-clause BSD license, it can serve as a basis for various DEM projects in research and commercial application, and is suited also for educational purposes.

Several possible extensions of the present code are easy to recognize. While a generalization to non-spherical particles (for example through rigid aggregates of spheres \cite{Bell:2005} or other representations \cite{Zhao:2023}) would require rather deep modifications to the code, extending the contact model by adhesion (typically with either the DMT or the JKR model, see \cite{Barthel:2008} for a review, and LAMMPS for an elegant implementation) is relatively straightforward. Other common DEM features that were left out here for simplicity are periodic boundary conditions, different classes of particles and mesh elements with different material properties, moving and rotating meshes, collisions with more primitive shapes (cylinders, cones, ellipsoids etc.), more flexible particle spawn methods, particle removal, different static and dynamic friction coefficients, a selection of constitutive models to choose from, a parameter file parser, etc.

On the numerical side, certain room exists for performance improvements. The present lean implementation is largely memory-bound, implying that it could likely benefit from parallelization for distributed memory systems with MPI. Moreover, more elaborate bookkeeping of nearby particle pairs---for example with Verlet lists \cite{Verlet:1967} rather than the naive reconstruction of linked cells in every timestep---could potentially speed up some simulations. For dilute or widely dispersed ensembles, binary spatial partitioning with a tree data structure will be more memory efficient than the regular grid used here, at the cost of a more involved implementation. Combined with mesh-aligned bounding boxes or voxelization of large slanted mesh elements, collision detection could potentially be sped up considerably. Such extensions are, however, beyond the scope of this work.

\subsection*{Acknowledgements}

Part of this work was funded by ETH Z\"{u}rich through ETHIIRA Grant no.\ ETH-03 10-3.

\subsection*{Competing Interests}

The author declares that there are no competing interests.

\end{document}